\documentstyle[12pt]{article}
\begin{document}

\title{\bf H-dibaryons and Primordial Nucleosynthesis} 

\author{ J.A.de Freitas Pacheco$^1$, S. Stoica$^{1,2}$,
 F. Th\'evenin$^1$ and J.E. Horvath$^{1,3}$ \\
\small {$^1$} Observatoire de la C\^ote d'Azur\\
 P.O. Box 4229, F-06304 Nice Cedex 4, France\\
\small $^2$ Institute of Physics and Nuclear Engineering \\
 P.O. Box MG-6, 76900-Bucharest, Romania\\
\small $^3$ Instituto Astron\^omico e Geof\'isico \\
 Av. Miguel Stefano 4200, S. Paulo  01060-970, Brazil }

\maketitle

\begin{abstract}

The apparent discrepancy between
abundances of light nuclides predicted by the standard Big-Bang 
and observational data is explained, by assuming the presence of  
metastable H dibaryons at the nucleosynthesis era. 
These dibaryons could be formed out of a small fraction of strange
quarks at the moment of the confinement transition.
For a primordial deuterium abundance of the order of
$3 \, \times \, 10^{-5}$, the measured differences in the $^4$He abundances
requires a relative abundance of H dibaryons
of the order of $n_H/n_B \sim $ 0.07, decaying in a
timescale of the order of 10$^5$ s.
\end{abstract}

Pacs : 14.20.Pt -Dibaryons, 98.80.Ft -Origin and formation of the elements

98.80.Es -Observational Cosmology

The abundances of D and $^3$He in the solar system and in the interstellar 
medium (ISM) \cite{[mol98]}- \cite{[Ste98]}, of  $^7$Li in metal-poor 
halo stars \cite{[SPI82]}, as well as of  $^4$He in low-metallicity 
extragalactic
HII regions \cite{[Ste98]} have been until recently in quantitative 
agreement with the standard Big-Bang nucleosynthesis (SBBN) predictions.        
However, recent observations reveal a 
possible conflict between the abundances of 
the above species predicted by SBBN and those inferred from observational data.
This situation has been generated by the new determinations of the deuterium 
abundance at high red-shift, low metallicity quasar (QSO) absorption clouds.
On the one hand, there is evidence in favor of a high value of the D abundance
(high-D), $(D/H) \sim 2\times 10^{-4}$  and, on 
the other hand, there is evidence for a low D abundance (low-D),
$(D/H) \sim 2.3\times 10^{-5}$. The former was derived from Keck data on
QSO 0014+813 \cite{[Son94]},  and it seems to be confirmed by new
data with better Signal to Noise ratio \cite{[Ty97]}, whilst the latter was
obtained
from the analysis of two QSO spectra \cite{[BT96]}, \cite{[Ty96]},
\cite{[BUR98]}. At present, no convincing arguments were given
in favor of one or other value, or even if these quite different
measurements indicate a non-homogeneous distribution.

From the SBBN, high-D 
values are consistent with  primordial abundances of  $^4$He 
and $^7$Li provided the baryon-to-photon ratio $\eta$ is about  
$3 \, \times \, 10^{-10}$
and three neutrino flavors  \cite{[Ste98]}. 
Note that low-D values imply a 
conflict between the predicted $^4$He abundance and observations.
More precisely, to have a consistency with SBBN, low-D values
require $^4$He abundances (by mass) in the range 0.247--0.250, whereas
extragalactic HII regions data indicate a primordial abundance of 0.230 $\pm$
0.003 \cite{[Ste98]}.

Several explanations for this apparent conflict have been proposed, including
corrections to the HeI effective line recombination coefficient,
uncertainties in the astration amount (model-dependent), and 
modification of the effective neutrino number. Basically,
the  predicted $^4$He abundance $Y_P$ depends
on the number of neutrinos with variations of the value
as $\Delta Y_P \sim 0.01(N_{\nu}-3$). SBBN is 
based on the hypothesis of three neutrino flavors ($N_{\nu}$=3). However, if 
$N_{\nu}$ = 2,  this will lead to negative correction and thus, a smaller 
value of $Y_P$, which could restore the consistency. 
An argument for a smaller $N_{\nu}$ would be the existence of a 
massive $\nu_{\tau}$.
For instance, a $\nu_{\tau}$
which would decay  with a 
lifetime of $\sim 0.1\, s$, reduces respectively
$N_{\nu}$ by $\sim 0.5$--$1$ and $Y_P$ by $\sim
0.006$--$0.013$ if its mass is 20--30 MeV and
solves the apparent conflict between 
theory and observations \cite{[HAT96]}. But,
the most recent estimation of the upper limit of
the mass of $\nu_{\tau}$ (18.2 MeV) weakens this argument \cite{[CAS98]}.

In this Letter we propose an alternative solution for this problem, 
by assuming the presence of  heavy metastable 
hadrons at the nucleosynthesis
era. These particles will eventually decay into protons, modifying the original
relative abundances. The presence of  massive and unstable particles X, formed 
just after the quark-hadron phase transition (QHT),
and with lifetimes longer than the nucleosynthesis era, was
also recently claimed as a possible mechanism to modify
the original SBBN abundances \cite{[Dim98]}. However, in such a scenario,
these unknown
particles decay and give rise to both electromagnetic and hadronic cascades.
The resulting high-energy photons would disintegrate a fraction of 
the original $^4$He, whereas the high-energy hadrons would
produce light nuclides by spallation reactions.

Our scenario supposes the presence of a small
fraction of  s-quarks at the moment of the QHT, 
which occurred when the temperature of the quark-gluon plasma
was about $T \, \sim \, 100-150 \, MeV$. In our picture, 
the strange quarks will confine
into dibaryons (H particles) \cite{[JAF77]},
after the QHT. The fraction of dibaryons just after
the QHT is determined by the density of s-quarks,  conservation
of electric and baryonic charge as well as strangeness. Most of
baryons will be annihilated at $T \, \sim \, 40 \, MeV$ and, 
as a consequence, the
baryonic charge excess is the most important thermodynamic parameter, implying
non-zero chemical potentials for quarks. The {\it u} and {d} quark flavor
excess density,
considering only first order terms in the chemical potential is ($\hbar$=c=1)
\cite{[SHU]}
\begin{equation}
n_f = 2T^3(1 - {{2\alpha_c}\over{\pi}}){{\mu_f}\over{T}},
\end{equation}
where $\alpha_c$  is the strong color coupling constant, and we have
assumed zero rest masses for these light quarks. Since the mass of the
{\it s} quark is still uncertain, its flavor density
excess is defined with respect to the {\it u} and {d} density by
$n_s = f_s(n_u + n_d)$. 

The baryon excess is $\varepsilon_B = {1\over 3}\sum{{n_f}\over{n_{\gamma}}}$,
where
$n_{\gamma}$ is the photon density. By using (1), this excess reads
\begin{equation}
\varepsilon_B = {{\pi^2}\over{3\zeta (3)}}(1 + f_s)({{\mu_u}\over{T}} + 
{{\mu_d}\over{T}})(1 - {{2\alpha_c}\over{\pi}}).
\end{equation}
The baryon excess must be followed by a lepton excess, which will be assumed
here to be carried only by electrons.
Electrical charge neutrality in the quark phase implies that
\begin{equation}
\lbrack(2 - f_s){{\mu_u}\over{T}} - (1 - f_s){{\mu_d}\over{T}}\rbrack(1 -
{{2\alpha_c}\over{\pi}}) = {1\over 2}{{\mu_e}\over{T}},
\end{equation} 
where $\mu_e$ is the electron chemical potential associated to
the lepton excess.

The H-dibaryon excess density with respect to the photon density is fixed
by strangeness conservation, i.e.,
\begin{equation}
{{n_H}\over{n_{\gamma}}} = {{\pi^2}\over{2\zeta (3)}}f_s(1 - {{2\alpha_c}
\over{\pi}})({{\mu_u}\over{T}} + {{\mu_d}\over{T}}).
\end{equation}
The excess of baryonic charge and electrical neutrality
just after the QHT give respectively the conditions
\begin{equation}
{{n_p}\over{n_{\gamma}}} + {{n_n}\over{n_{\gamma}}} 
+ {{2n_H}\over{n_{\gamma}}} = \varepsilon_B,
\end{equation}
and
\begin{equation}
{{n_p}\over{n_{\gamma}}} = {{\pi^2}\over{6\zeta (3)}}{{\mu_e}\over{T}},
\end{equation}
where $n_p$ and $n_n$ are respectively the proton and the neutron density 
excess. This system of equations determines the initial abundance of dibaryons 
as a function of the parameter $f_s$, which will 
mostly contribute to the appearance of H particles after hadronization.
After decay of H's, the dilution of helium can be inferred and 
compared to the measured abundances. 

An initial H fraction of about 7\% is sufficient to explain the apparent $^4$He 
discrepancy,
without modifying considerably the abundances of the other nuclides. We now
address the crucial question regarding the ability of H dibaryon
to survive until the nucleosynthesis era ($T \, \sim \, 1 \, MeV$).
We recall that the existence of the H-hadron was
first suggested by Jaffe \cite{[JAF77]} in 1977,
as a double strange ($\Lambda \Lambda$), flavor-singlet 
six-quark state (uuddss), with spin and parity $J^{\pi}=0^+$ and isospin 
$I=0$, which would be stable against strong decay.  
Concerning the H dibaryon mass,  theoretical estimations 
\cite{[JAF85]}-\cite{[TWA88]} vary from a very bound state with respect to 
the $\Lambda \Lambda $ threshold, i.e. $\simeq 2231$ MeV, (within a quark 
cluster model) \cite{[SAK97]} to a slightly bound state
(within the chiral quark model) \cite{[BAL84]},
or even an unbound state (from a lattice QCD 
calculation) \cite{[MAC85]}. The latter case seems unrealistic.
Experiments are not yet decisive   
\cite{[IMA91]}-\cite{[AHN96]}, and speculations on the dibaryon 
mass are still possible.  

If the H dibaryon mass is smaller than the $\Lambda\Lambda$-threshold, then it 
does not decay by strong interactions. Thus, its lifetime is not 
related to the 
lifetime of free $\Lambda$ hyperons (which is of the order of $\sim 
10^{-10}$ s).  If  the H dibaryon is regarded as a six-quark composite 
particle, described by a wave function involving the symmetry properties of 
the ground state configuration,  the picture of its 
weak decay and lifetime will be different from that of a two-hyperon decay.
In this case H can decay through the nonleptonic modes 
$\Delta S= 1, 2$ (S is the 
strangeness quantum number). If $m_{\Lambda} + m_n < m_{H} < 2 m_{\Lambda}$, 
the H decays through the $\Delta S = 1$ mode into the following three 
channels : $n\Lambda$, $n\Sigma^0$ and $p\Sigma^-$. A small 
contribution can also be the
$\Lambda N \pi$ channel. The lifetime
for this channel  has been calculated in detail in Ref. \cite{[DON86]}. 
The authors first build up a six-quark wave functions and then  
construct the $\Delta S= 1, 2$ nonleptonic weak 
Hamiltonians and compute the decay rates. They have found for this mode 
lifetimes between $3\times 
10^{-9}- 7 \times 10^{-7}$, for a mass interval between $2.22$ and $2.06$ 
GeV. These lifetimes are too short to permit dibaryons to survive
during the nucleosynthesis era. 
However, if the H dibaryon has a mass below the 
$n + \Lambda$ threshold, i.e. $m_H \leq m_n + m_{\Lambda}$ (i.e. $m_H \leq 
2054$ 
MeV), all $\Delta S= 1$ decay channels are forbidden and they can  
decay only by the $\Delta S =2 $ mode. In this case, the expected lifetimes are
of the order of a few days ($\sim 10^5$ s) \cite{[DON86]}, 
sufficient for their survival 
during the entire period of formation of the light nuclides. 

Another important question concerns the survival of dibaryons against
collisions with other more abundant baryons. If their relative abundance
with respect to photons is $X_H = {{n_H}\over{n_{\gamma}}}$, then
their evolution is
governed by the equation
${{\partial X_H}\over{\partial t}} = -X_H\nu_i$
where $\nu_i $ is the collision frequency with other nucleons 
(protons and neutrons).
The relevant energy scale is the difference 
between the actual H mass  and the minimum 
energy that would make the excited dibaryon to decay on a timescale 
$\tau \, \ll \, 10^{5} \, s$. Even if the typical thermal 
energies of the nucleons stay always less than the difference 
$\Delta \geq 175 MeV$ between two lambdas and the H mass,  
this does not guarantee the survival of a sufficient number of H's 
until the nucleosynthesis era. If the QHT occurs at about ${1\over{H_H(t)}}
\approx$ 10 $\mu$s ($H_H(t)$ being the Hubble parameter), then the
inelastic cross section must satisfy $\sigma <$ 10$^{-10}$ barn, to avoid
a significant destruction. 

Finally it is worth mentioning that H decays into two neutrons with energies 
of the order of hundred MeV, and these in turn decay into two protons, 
two electrons and two anti-neutrinos. Such processes contribute to
a heat input in the primordial plasma, but with negligible consequences.
H-dibaryons with a lifetime of about 10$^5$ s will decay when the
primordial plasma reaches a temperature of about 1.5 keV, corresponding
to an energy density of about 7$\times 10^{14}$ erg.cm$^{-3}$. The
H decay implies an energy input of about 1.4$\times 10^9$ erg.cm$^{-3}$,
which will introduce a quite small correction into the plasma temperature.
Then the possible presence of strange hadrons (H-dibaryons)
able to survive from the QHT until the nucleosynthesis epoch in the
early universe, could be an alternative solution keeping compatible
the predictions of the SBBN and observational data. Our scenario
would implies $\eta \approx \, 5 \, \times \, 10^{-10}$, or equivalently
$\Omega_b \approx$ 0.04 (H$_0$ = 65 km/s/Mpc).

\end{document}